\documentclass{ifacconf}

\usepackage{graphicx}      

\usepackage{natbib}        

\usepackage{amsmath,amsfonts}

\usepackage{array}
\usepackage{textcomp}
\usepackage{stfloats}
\usepackage{url}
\usepackage{verbatim}
\usepackage{balance}
\usepackage{xcolor}
\usepackage{lineno}
\modulolinenumbers[5]
\usepackage{siunitx}
\usepackage{algorithm}

\usepackage{algpseudocode}

\DeclareMathOperator{\euc}{EUC}
\DeclareMathOperator{\dtw}{DTW}
\DeclareMathOperator{\dist}{dist}

\DeclareMathOperator{\simi}{\mathcal{}S}

\begin{document}
\begin{frontmatter}

\title{A Two-Stage Machine Learning-Aided Approach for Quench Identification at the European XFEL}

\thanks[footnoteack]{The authors acknowledge support from DESY (Hamburg, Germany), a member of the Helmholtz Association HGF. This work was funded in the context of the R\&D program of the European XFEL. The authors would like to thank Jan Horst Karl Timm, Christian Schmidt and Nicholas Walker for their help.}

\author[First]{L. Boukela} 
\author[First,second]{A. Eichler} 
\author[First]{J. Branlard}
\author[First]{N.Z. Jomhari}

\address[First]{Deutsches Elektronen-Synchrotron DESY, 22607 Hamburg, Germany (e-mail: lynda.boukela@desy.de, annika.eichler@desy.de, julien.branlard@desy.de, nur.zulaiha.jomhari@desy.de).}

\address[second]{Hamburg University of Technology, 21073 Hamburg, Germany (e-mail: annika.eichler@tuhh.de).}

\begin{abstract}      
This paper introduces a machine learning-aided fault detection and isolation method applied to the case study of quench identification at the European X-Ray Free-Electron Laser. The plant utilizes 800 superconducting radio-frequency cavities in order to accelerate electron bunches to high energies of up to \SI{17.5}{GeV}. Various faulty events can disrupt the nominal functioning of the accelerator, including quenches that can lead to a loss of the superconductivity of the cavities and the interruption of their operation. In this context, our solution consists in analyzing signals reflecting the dynamics of the cavities in a two-stage approach. (I) Fault detection that uses analytical redundancy to process the data and generate a residual. The evaluation of the residual through the generalized likelihood ratio allows detecting the faulty behaviors. (II) Fault isolation which involves the distinction of the quenches from the other faults. To this end, we proceed with a data-driven model of the k-medoids algorithm that explores different similarity measures, namely, the Euclidean and the dynamic time warping. Finally, we evaluate the new method and compare it to the currently deployed quench detection system, the results show the improved performance achieved by our method.
\end{abstract}


\end{frontmatter}

\section{Introduction}
Fault detection plays a crucial role in ensuring the
safe and optimal operation of complex systems. Traditional model-based approaches, although widely explored, have some limitations in their ability to isolate all possible faults, especially those with evolving patterns as complete models are usually hard to obtain. Machine learning (ML) approaches, on the other hand, are simpler and handle more effectively complex systems, however, their performance relies heavily on the data quantity and quality. Hybrid solutions seem to be promising although the choice of methods to integrate is challenging 
 \citep{wilhelm2021overview}. A hybrid method is thus developed for our case study, i.e., the problem of quench detection at the European X-Ray Free-Electron Laser (EuXFEL), where a fault detection method based on physical model residuals and data-driven clustering for fault isolation are combined.

The EuXFEL is the largest particle accelerator for X-ray laser generation worldwide. Along with the high-voltage power supplies, the klystrons and the waveguides, the superconducting radio-frequency cavities (SRFCs), are key components of the power transfer chain to the electron beam. The plant's linac comprises hundreds of \SI{1.3}{GHz} TESLA-type cavities, controlled and monitored with the low level radio frequency (LLRF) system and operated in a \textit{pulsed} mode with a pulse repetition rate of \SI{10}{Hz} and at an average gradient of \SI{23.6}{MV/m} \citep{reschke2017commissioning}. This enables the three self-amplified spontaneous emission undulators to generate up to \SI{27000} photon flashes every second. This setup is exploited by several hundred interdisciplinary users every year to carry out their experiments.

Enhancing the availability, safety and reliability of the accelerator is therefore crucial to meet the demand, minimize potential energy and financial losses and avoid the degradation of the facility. In this context, the behavior of the SRFCs is monitored through the LLRF system to detect and report the faulty events, especially quenches. These are severe faults that cause a loss of the superconductivity of the SRFCs and thus an operation down-time. If a quench goes undetected, it has the potential to generate sufficient heat that can lead to disturbances in the helium flow of the cryogenic system and to an operation interruption that could last up to 24 hours.

A quench detection system (QDS) has been deployed at the EuXFEL since its first commissioning in 2017 \citep{ayvazyan2013superconducting}. It relies on a statistical analysis of the pulses for indication of changes of the cavity quality factor $Q_{L}$, which is an indicating measure of the field coupling and the dissipating power in the SRFCs. The QDS is, however, not robust enough, as disruptive events such as controlled detuning, field emitters, and digital glitches can affect the $Q_{L}$ computation, and lead the system to trigger alarms as a result of \textit{fake} quenches \citep{eichler2023anomaly}. Because of the considerable number of these false alarms, and since the physical model of the SRFCs and measurements from the LLRF system are available to build a more robust quench detection method, alternative solutions have been explored and developed \citep{Nawaz2016, nawaz2018anomaly, Syed2021, Bellandi2021, Sulc2022, Martino2022, Branlard2022ML, eichler2023anomaly}. Purely data-driven techniques have also been investigated in other facilities \citep{tennant2020superconducting, vidyaratne2022deep}.

In our solution, the well-established model of the nominal behavior of the SRFCs from \citep{schilcher1998vector} is exploited in order to obtain a residual. To this purpose, the parity space-based approach has been retained for fault detection \citep{eichler2023anomaly}, followed by the main contribution of this paper, the ML-assisted quench isolation. The main motive for the usage of the parity space is the robustness of the method against drifts and noise, but also due to the need of a firmware implementation. A statistical test of the residual, namely the generalized likelihood ratio (GLR), that presents a distribution that alleviates the process of threshold determination, is used to detect all occurring faults. The test outputs fault-specific patterns, and to further distinguish the quenches from other types of faults, the k-medoids clustering algorithm is employed to learn a quench model. Two clustering models are obtained with two different similarity measures, namely, the lockstep Euclidean (EUC) and the elastic dynamic time warping (DTW). The inference is achieved through thresholds and decision boundaries on the distances to the quench medoids. This approach presents advantages in terms of suitability for an online processing setup and with respect to interpretability which is an important aspect for the LLRF experts.

Section \ref{S1} presents the model-based residual generation and evaluation for fault detection. Section \ref{S2} details the quench identification through the data-driven fault isolation. The experimental evaluation setup and results are discussed in Section \ref{S3}. Finally, Section \ref{S4} concludes the paper.

\section{Residual-based fault detection}\label{S1}
Model-based fault detection relies on the assessment of consistency between available measurements and prior knowledge. This is achieved through residuals that can be obtained with different methods. In our case, the nonlinear parity space method is used, and it is detailed here.

\subsection{The physical model of the SRFCs}
Because of its structure that is able to exhibit a resonant behavior at specific frequencies, a superconducting cavity is considered as a resonator. The parameters influencing the cavity's dynamics are the detuning $\Delta \omega(t)$ and the half bandwidth $\omega_{1 / 2}$ which represent, respectively, the difference between the driving frequency and the resonance frequency, and how sensitive a SRFC is towards the detuning. The cavity electromagnetic model as elaborated in \citep{schilcher1998vector} is given by 
\begin{equation}\label{model_eq}
\begin{aligned}
{\left[\begin{array}{c}
\dot{V}_{P, I}(t) \\
\dot{V}_{P, Q}(t)
\end{array}\right]=} & {\left[\begin{array}{cc}
-\omega_{1 / 2} & -\Delta \omega(t) \\
\Delta \omega(t) & -\omega_{1 / 2}
\end{array}\right]\left[\begin{array}{c}
V_{P, I}(t) \\
V_{P, Q}(t)
\end{array}\right] } \\
& +2 \omega_{1 / 2}\left[\begin{array}{c}
V_{F, I}(t) \\
V_{F, Q}(t)
\end{array}\right]-\omega_{1 / 2}\left[\begin{array}{c}
V_{B, I}(t) \\
V_{B, Q}(t)
\end{array}\right],
\end{aligned}
\end{equation}

where I and Q are the in-phase and quadrature components of the different signals. $V_{F}(t)$ is the forward field coupled into the cavity, ${V}_{P}(t)$ is the probe signal, i.e., the field generated inside the cavity, and $V_{B}(t)$ is the field induced by the beam. 

\subsection{Residual generation}
The nonlinear parity space is an analytical redundancy method that aims at generating redundant expressions using the state-space model. Successive derivations of the model must be performed in order to retain only the known variables and obtain residuals. The residuals are then used as indicators of faults presence, a residual is non-zero in presence of faults, it is equal to zero otherwise. For the problem at hand, the cavity model has been exploited to derive the residual 
\begin{equation*}\label{re_eq}
\begin{aligned}
r(t)= & \frac{-\dot{V}_{P, I}(t)+\omega_{1 / 2}\left[-V_{P, I}(t)+2 V_{F, I}(t)-V_{B, I}(t)\right]}{V_{P, Q}(t)} \\
& -\frac{\dot{V}_{P, Q}(t)+\omega_{1 / 2}\left[V_{P, Q}(t)-2 V_{F, Q}(t)+V_{B, Q}(t)\right]}{V_{P, I}(t)},
\end{aligned}
\end{equation*}

 that is based on the analytical redundancy of (\ref{model_eq}) with respect to the detuning. More details regarding the choice of the residual can be found in \citep{nawaz2018anomaly}.
 
\subsection{Residual assessment}
In order to automate the residual analysis for presence of faults, the log-likelihood ratio is used with the assumption that the residual follows a Gaussian distribution $\mathcal{N}(\mu, \sigma)$ with a mean of zero under nominal conditions, i.e., $\mu = 0$, representing the null hypothesis. While with fault occurrence, the residual is expected to exhibit a Gaussian distribution with a mean different from zero, i.e, $\mu \neq 0$, representing the non-null hypothesis. Unlike the variance, the mean is unknown and needs to be estimated, the generalized likelihood ratio is therefore applied, 
\begin{equation*}\label{glr_eq}
\begin{aligned}
\lambda_{\mathrm{GLR}}(k)\!=\!\frac{K}{2}\left(\frac{1}{K}\!\sum_{i=k-K+1}^k\!r(i)^{\top}\right)\!\sigma^{-1}\left(\frac{1}{K} \!\sum_{i=k-K+1}^k\!r(i)\right),
\end{aligned}
\end{equation*} 

where $r(i)$ is the evaluated and discretized residual, $K$ is the size of the moving evaluation window and $\sigma$ is the variance of the nominal residual. 
The result follows a ${\chi}^2$ distribution, which gives, with a desired false alarm rate, directly the corresponding threshold on rating events as anomalous \citep{eichler2023anomaly}.

\section{Clustering-based quench isolation}\label{S2}
The statistical test helps to detect faults, it is however unable to identify their types. 
Data-driven methods help to address this issue. 
The goal in our case is to isolate the quenches from the rest of faulty data. The k-medoids algorithm \citep{kaufman2009finding} is thus used to obtain quench clustering models from the statistical results. Furthermore, decision boundaries are used in addition to thresholds in order to ensure an optimal separation of the different faults during the inference.

\subsection{Quench clustering models}
The k-medoid method is briefly summarized in the following. Given  $X=\{x_1, x_2, \ldots, x_n\}$, with $x_i\in\mathbb{R}^{d}$, a dataset of $n$ observations with dimension $d$. The k-medoids is a clustering algorithm that aims to partition $X$ into $k$ clusters $ C=\{c_{1}, c_{2},\ldots, c_{k}\}$ by iteratively assigning every observation to a cluster medoid $M=\{m_{1}, m_{2},\ldots, m_{k}\}$, here, a medoid is a representative observation, i.e., $M \subset X$. In the same cluster, observations show high degree of similarity $\simi(x_1, x_2)$, while they are as dissimilar as possible in different clusters. The clustering can be achieved with different partitioning algorithms. The main input parameters are the number of clusters $k$, the similarity measure $\simi$ and the maximum number of iterations. 

\begin{figure}
\begin{center}
\includegraphics[width=8.8cm]{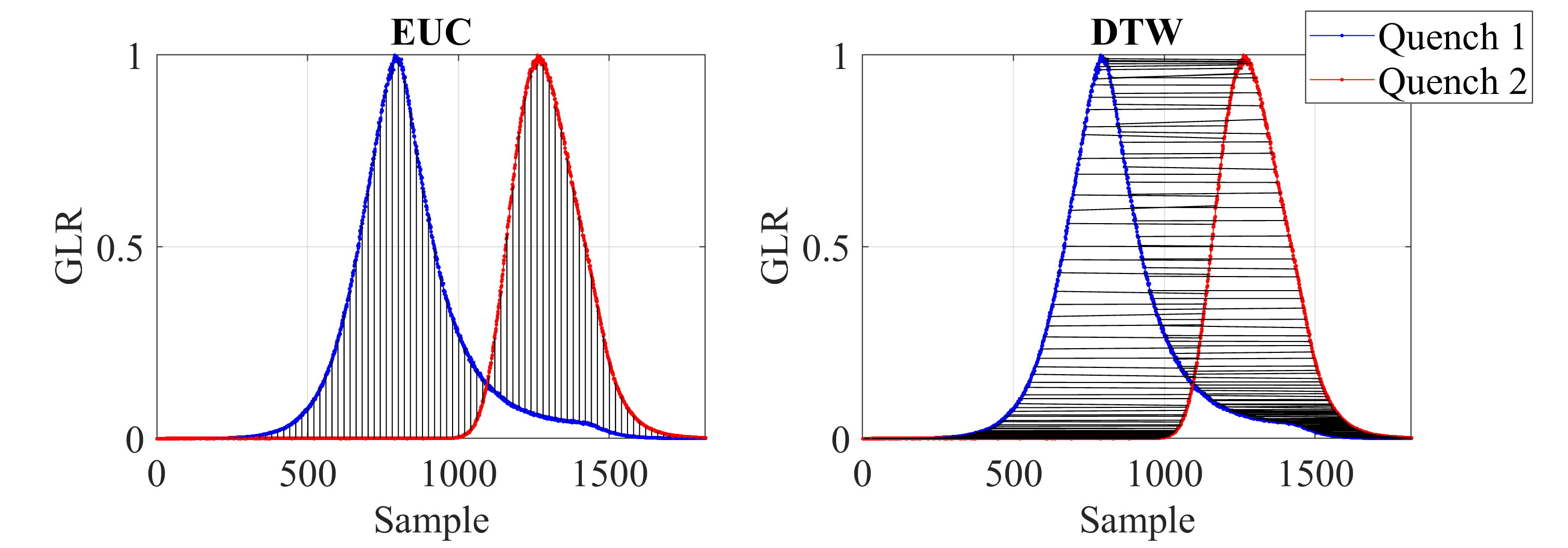} 
\caption{Illustrative example of the DTW and EUC similarity measures, and their difference in terms of sample correspondence. Here, DTW is equal to 2.8 and EUC is equal to 18.6.} 
\label{dtw_euc}
\end{center}
\end{figure}

Two models are built in our case, based on two different similarity measures. As base measure, the EUC is used,
\begin{equation*}
\euc(x_1,x_2) = \sqrt{\sum_{i=1}^{n}(x_1^i-x_2^i)^2}.
\end{equation*}
Given the GLR traces of normalized quench pulses that exhibit variability, characterized primarily by stretches and shifts in time, as shown in Fig. \ref{dtw_euc}, the DTW 
 \citep{sakoe1978dynamic}
\begin{equation*}
\dtw(x_1,x_2) = \arg\min_{i, j}{\sum \dist(x_1^i, x_2^j)},
\end{equation*} 

is also explored, where $i, j \in \{1, 2, \ldots, d\}$ are sample indices of the observations $x_1$ and $x_2$. Any distance ($\dist$) can be employed for DTW, here, the Euclidean is used.

The difference between the two, as can be seen from the previous equations and from the illustrative example in Fig. \ref{dtw_euc}, consists mainly in what the measures are trying to capture. While the EUC computes the similarity as the sum of the distances between each sample from observation $x_1$ and its time-corresponding sample in observation $x_2$, DTW is the sum of distances from the optimal time-warping path, i.e., the one assigning samples of observation $x_1$ to the closest samples in observation $x_2$. DTW tries therefore to capture the shape similarity in addition to the distance. And the quench traces in Fig. \ref{dtw_euc} are thus more similar with respect to DTW that is equal to 2.8 compared to EUC found to be equal to 18.6. Data from 2021, consisting of $n\,=\,76$ quench traces of dimension $d=1819$, have been used to build the models. The data are first pre-processed, a maximum-based normalization is adopted in order to equalize the scale of the data without changing its inherent structure that is important for DTW. The scaling is further accompanied by a reduction, in order to retain the frame of interest from the traces. It is obtained by keeping 300 samples following the first sample greater than 0.2. This applies only to the Euclidean distance, as it is unable to capture the similarity of traces with time misalignment. 

 \begin{figure}
\begin{center}
\includegraphics[width=8.8cm]{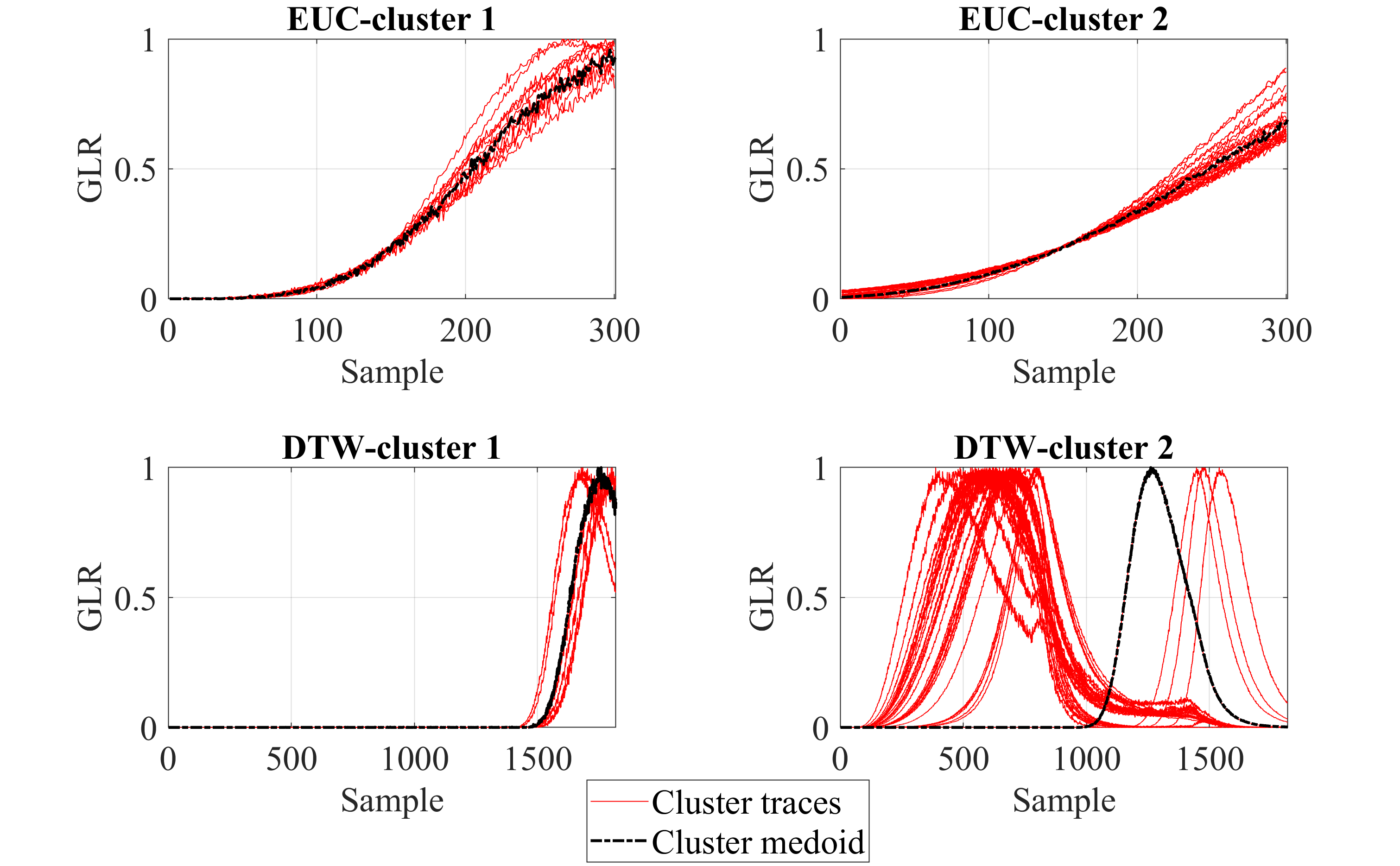} 
\caption{Traces and medoids of the two clusters obtained with the two distances, EUC and DTW.} 
\label{qtraces}
\end{center}
\end{figure}

 The models are then built using $k$-medoids. The Partitioning Around Medoids \citep{kaufman2009finding} algorithm is utilized. Two main patterns of the quench traces have been noticed, two clusters are therefore built, i.e., $k=2$, and two medoids $M = \{m_1, m_2\}$ are obtained with $100$ iterations. The traces and medoids of the clusters obtained with the two distances are shown in Fig. \ref{qtraces}.
 With both measures, the second cluster gathers the quenches appearing at an early stage, (at the beginning of the pulse), 
 while those in the first cluster are quenches that have occurred at a later stage (at the end of the pulse). 
 \subsection{ Inference }
Given a new faulty trace $g\in\mathbb{R}^{d}$, in order to make decision whether it is a quench or not, its distance to the quench medoids is computed to assess its similarity with the quench clustering model. For both similarity measures, we define a set, based on the distances to the medoids, within which a trace is labeled as a quench. To estimate the sets, we set lower and/or upper thresholds on the distance to the quench medoids, in addition to decision boundaries $\mathcal{H}$. To estimate $\mathcal{H}$, a validation dataset $N$, with a total of 407 non-quench faulty traces, is used. 
\subsubsection{EUC-based inference:}
\begin{figure}
\begin{center}
\includegraphics[width=8.8cm]{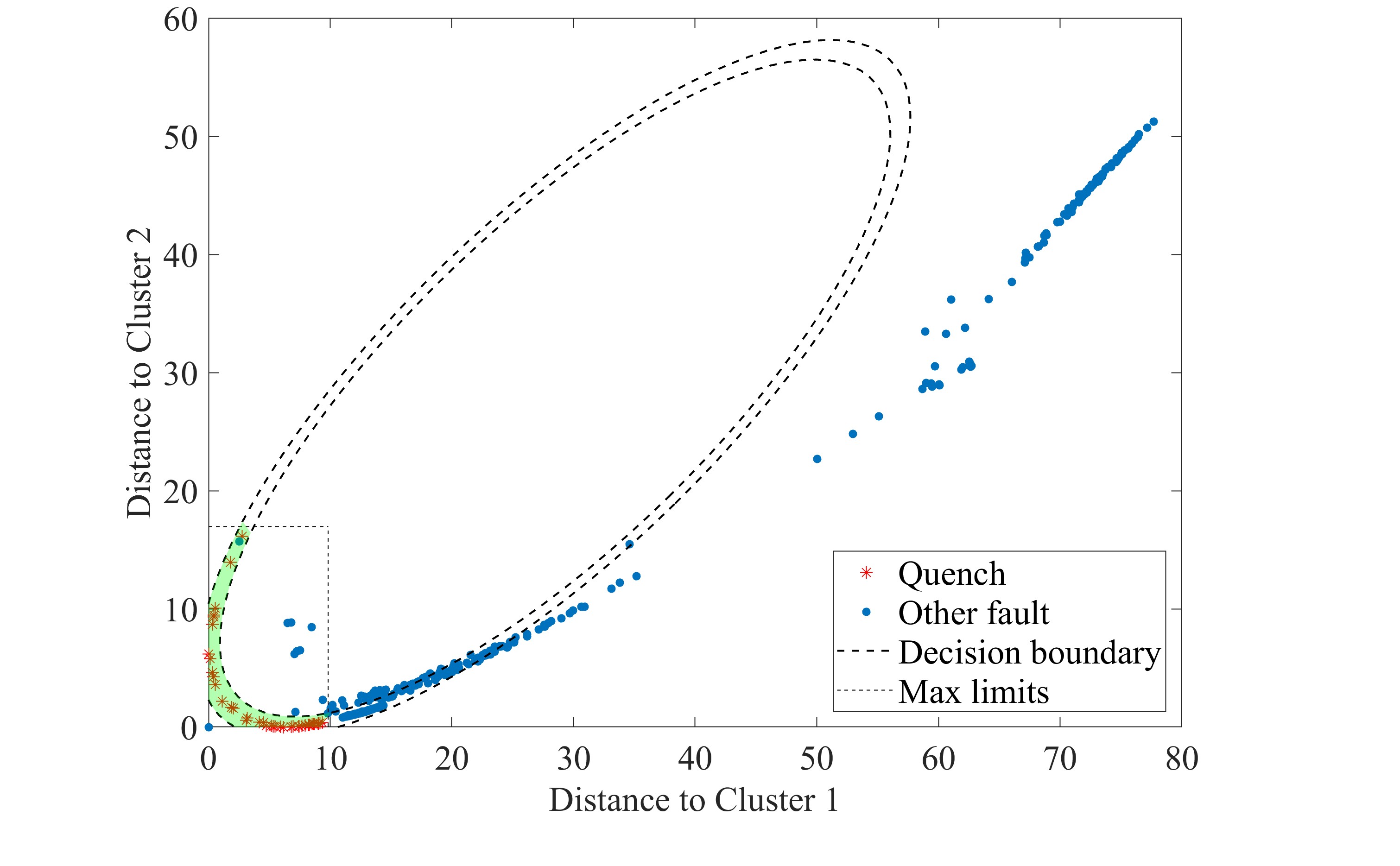}    
\caption{EUC decision boundaries ($\mathcal{H}_{EUC}^i$ and $\mathcal{H}_{EUC}^o$) and thresholds. The green delimits the area of the quench distances. 
} 
\label{fig4}
\end{center}
\end{figure}

Fig. \ref{fig4} depicts the distance space (distance from the medoids), obtained with EUC, of the two quench clusters from $X$ and the validation data $N$. As can be seen, the two quench clusters are not distinct and present a semi-elliptical shape. In order to separate the quenches from the other faults, and given the shape of the quench distances, in addition to the thresholds, two ellipses are used as decision boundaries, an inner boundary $\mathcal{H}_{EUC}^{i}$ and an outer boundary $\mathcal{H}_{EUC}^{o}$. The two ellipses are obtained as a scaling of the main ellipse $\mathcal{H}_{EUC}$ fitted to the quench distances through the Least Squares method, and satisfying the following equation
\begin{equation*}\label{EUCeq}
\begin{aligned}
\mathcal{H}_{EUC}(s_1, s_2)=\frac{\left(\left(s_1-s_1^{0}\right)\cos\left(\phi\right)-\left(s_2-s_2^{0}\right)\sin\left(\phi\right)\right)^{2}}{a^{2}}+ \\
\frac{\left(\left(s_1-s_1^{0}\right)\sin\left(\phi\right)+\left(s_2-s_2^{0}\right)\left(\cos\left(\phi\right)\right)\right)^{2}}{b^{2}}=r^2,
 \end{aligned}
\end{equation*}

where $s_1$ is the distance to the first cluster, $s_2$ is the distance to the second cluster, $\phi$ is the  rotation angle, $s_1^{0}$ is the center at the horizontal axis, $s_2^0$ is the center at the vertical axis and $r$ is the scaling factor equal to 1. 

A set of quench traces based on EUC can be defined as
\begin{equation*} \label{EUCQuench}
\begin{aligned}
\mathcal{Q}_{EUC}=\{ q \, | \, &q \in  \mathbb{R}^d, \\
&\hspace{-0.7cm}\euc(q, m_j) \,<\, \max(\euc(x_i, m_j))+\epsilon, \\
&\hspace{-0.7cm}(r^{i})^2 < \mathcal{H}_{EUC}(\euc(q,m_1), \euc(q,m_2))<(r^{o})^2,\\
&\hspace{-0.7cm}\forall x_i \in X, \, \forall m_j \in M \},
\end{aligned}
\end{equation*}

where $\epsilon$ is a small tolerance value learned empirically, $r^i$ and $r^o$ are the radii of $\mathcal{H}_{EUC}^{i}$ and $\mathcal{H}_{EUC}^{o}$, respectively, obtained as the closest and furthest distances, $\mathcal{H}_{EUC}(\euc(x_i,m_1), \euc(x_i,m_2)) \text{ with } x_i \in X$, minus or plus $\epsilon$. The detection for a new trace $g$ is therefore achieved as in the following
\begin{equation*}
\text{Inference}_{EUC}(g) =
\begin{cases}
    \text{Quench,} & \text{if } \, g \in \mathcal{Q}_{EUC} \\
    \text{Other fault,} & \text{otherwise}.
\end{cases}
\end{equation*}

\begin{figure}[h]
\begin{center}
\includegraphics[width=8.8cm]{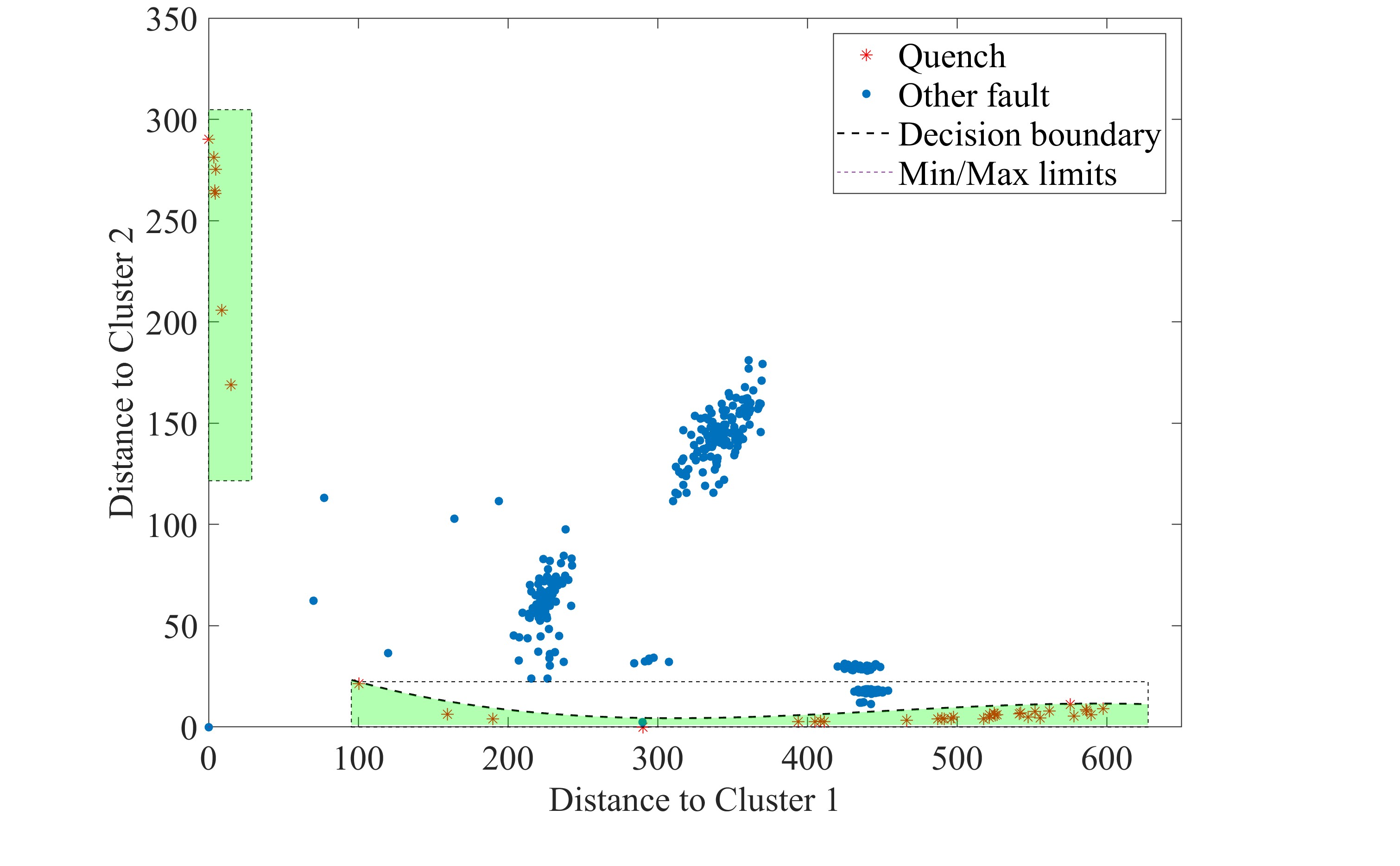}    
\caption{DTW decision boundary and thresholds. The green highlights the area of the quench distances.} 
\label{fig5}
\end{center}
\end{figure}

\subsubsection{DTW-based inference:}

Fig. \ref{fig5} depicts the distance space, obtained with DTW, of both the quench clusters from $X$ and the validation data $N$. Here, we can see a clear separation of the two trace types. While the quench cluster in the top left of the plot is well isolated from the non-quench faults, some overlapping can be noticed between the other faults and the second quench cluster, a decision boundary is therefore fitted as follows 
\begin{equation*} \label{DTWeq}
    \mathcal{H}_{DTW}(s_1)= as_1^3 + bs_1^2 +cs_1+ f, \\
\end{equation*}

with $a,\, b,\, c \text{ and } f$ being the polynomial coefficients. The decision boundary is obtained by adapting the intercept of the cubic fit of the distances in the cluster in the bottom of the plot. 

Let $\mathcal{Q}_{DTW}$ be the set of quench traces based on DTW 
\begin{equation*} \label{DTWQuench}
\begin{aligned}
\mathcal{Q}_{DTW}=\{ q \, | \, & q \in  \mathbb{R}^d, \\
&\hspace{-0.7cm}\dtw(q, m_j) \,< \max(\dtw(x_i, m_j))+\epsilon, \\
&\hspace{-0.7cm}\dtw(q, m_j) \,> \min(\dtw(x_i, m_j))-\epsilon, \\
&\hspace{-0.7cm}\dtw(q,m_2)<\mathcal{H}_{DTW}(\dtw(q, m_1)),\\
&\hspace{-0.7cm}\forall x_i \in X, \, \forall m_j \in M \},
\end{aligned}
\end{equation*}
where $\epsilon $ is a small tolerance value. The inference for a new trace $g$ is therefore achieved as follows
\begin{equation*}
\text{Inference}_{DTW}(g) =
\begin{cases}
    \text{Quench,} & \text{if } \, g \in \mathcal{Q}_{DTW} \\
    \text{Other fault,} & \text{otherwise}.
\end{cases}
\end{equation*}

\section{Evaluation}\label{S3}

\subsection{Setup}

The EuXFEL linac is organized into 25 stations, each comprising 4 cryomodules. Within each cryomodule, there are 8 cavities, resulting in a total of 32 SRFCs per station. The data from the SRFCs consist of the forward and probe signals that correspond to the radio frequency pulses, where each pulse lasts for \SI{1.82}{ms} and is sampled at 1 MHz (i.e., 1819 samples). The data are captured by a snapshot recorder and saved in files in the hdf5 format. The snapshots are triggered by different protection systems, such as, klystrons, modulators, couplers, quadrupoles and  cryogenics, but also the quench detection system and/or the finite state machine. Each file, identified by a unique timestamp, corresponds to a specific faulty event from an individual station, and gathers the sampled signals for 250 consecutive pulses in all the cavities of the station. Results of the residual-based detection, i.e., the GLR traces, are also saved to the same file. The ground truth for the specific faults, however, is not available. For this study, data from the first half and second half of 2022, corresponding to a total of 671 station-related snapshots, have been collected and annotated in order to evaluate the performance of the proposed approach. 

We compare the proposed solution to the currently deployed QDS that relies on a statistical analysis of the quality factor $Q_{L}$ of the cavities. The $Q_{L}$ is computed for almost every pulse and compared to a running average from the previous 100 pulses, and a sudden drop of the $Q_{L}$ is an indicator of a quench. Results of the QDS are event-wise, i.e., the QDS pulse-by-pulse predictions are not available. In our case, an event is considered as a quench if at least one pulse in the station is identified as a quench. This is applicable to the problem at hand, as the identification of a quench results in the interruption of the radio frequency driving the affected station. 

In order to illustrate the performance comparison in terms of quench identification, the area under the curve (AUC) of the receiver operating characteristics (ROC) curves is used. The $ROC\text{-}AUC$ depicts the true positive rate 
\begin{equation*}
TPR=\frac{TP}{(FN+TP)},
\end{equation*}
as a function of the false positive rate
\begin{equation*}
FPR=\frac{FP}{(FP+TN)}, 
\end{equation*}
where $TP$ are the true positives (i.e. the method accurately identified a quench), $TN$ are the true negatives (i.e. the method accurately recognized a fault was not a quench), $FP$ are false positives (i.e. the method identified a quench that's in reality another fault), and $FN$ are the false negatives (i.e., the algorithm failed to identify a real quench). 
 
\subsection{Implementation}
The online and offline residual generation and evaluation are integrated to the trip event logger \citep{timm2021trip}, an efficient tool implemented in C++ for automatic fault handling and prevention. The machine learning clustering models, along with an annotation tool, being at a testing phase so far, have been implemented in Matlab. The clustering models are deployed offline, and run daily to analyse the statistical results on the hdf5 snapshots.

\subsection{Results}

Results of the residual-based fault detection  have shown that given the total number of 671 events, 354 were detected as faults within the SRFCs, some of them are documented in \citep{eichler2023anomaly}. After annotation, 87 events were found corresponding to quenches. 
The highest number was recorded in April with a total of 21 quenches. This is because the facility was running at high energies, i.e., most cavities were operated just below the quench gradient. More detailed evaluation of the residual-based fault detection, and examples of the events that triggered alarms with the GLR, can be found in previous publications \citep{nawaz2018anomaly, eichler2023anomaly}. 

It is worth mentioning that the current residual implementation does not include the voltage induced by the beam $V_B$, as additional input, as shown in (\ref{model_eq}), but its effect is compensated as presented in \citep{eichler2023anomaly}. This becomes problematic in certain rare cases where inaccurate inferences may be triggered due to transitions of the beam turning on and off or changes in its pattern occurring within the 250 pulses captured in the snapshots. First tests where the bunch charge measurements from the toroids are added as an additional input have therefore been carried out, this will be automated and evaluated in the future.

\begin{figure}
\begin{center}
\includegraphics[width=8.8cm]{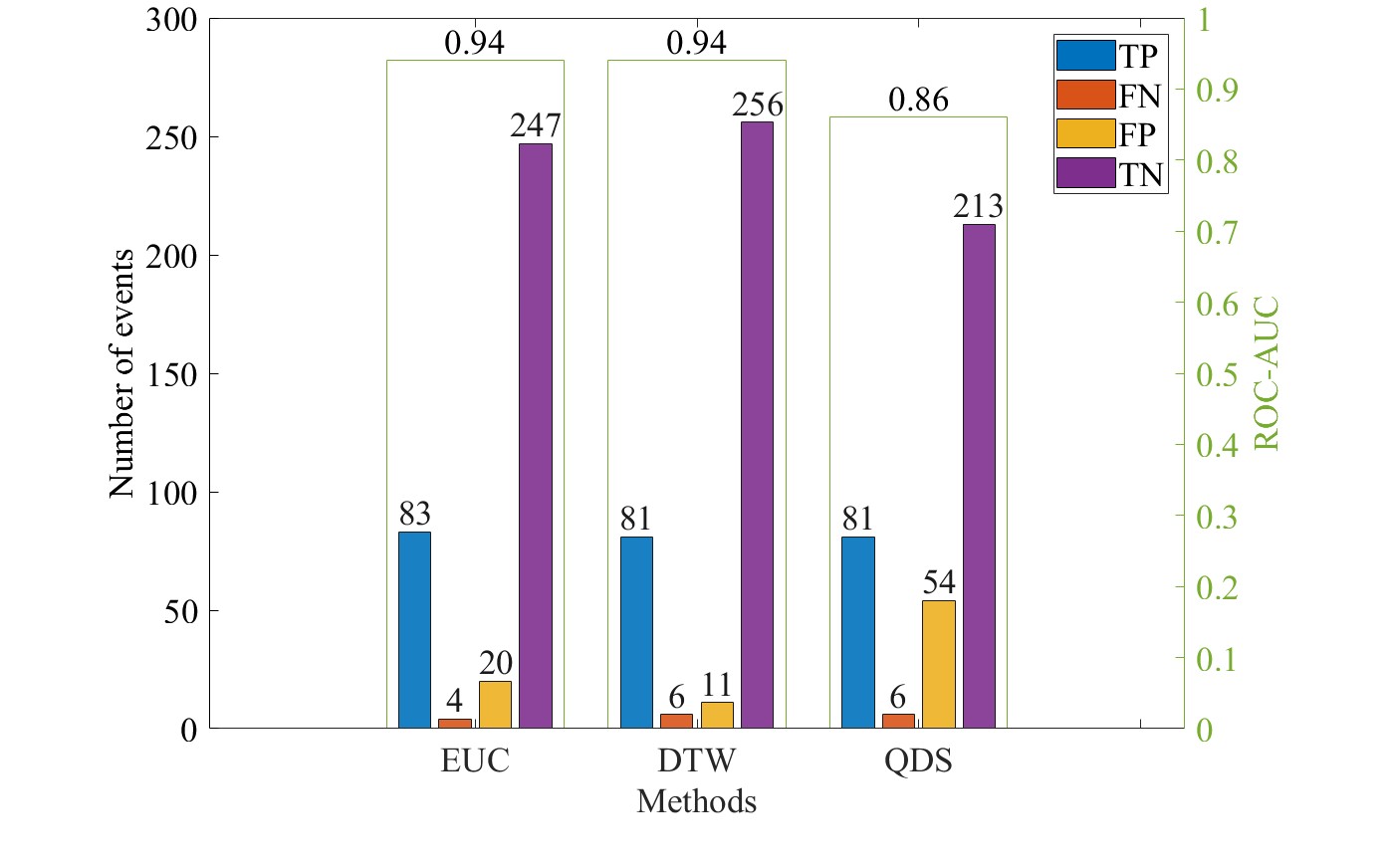} 
\caption{Comparison of the quench identification results obtained with the different methods.} 
\label{fig6}
\end{center}
\end{figure}

The evaluation is based on the traces of the 354 faulty events. Results, obtained with the different methods for quench isolation, namely, EUC, DTW and the currently deployed QDS, are presented in Fig.\ref{fig6}. The plot depicts the obtained number of $TP$, $FN$, $FP$, $TN$ and the $ROC\text{-}AUC$. The two clustering-based methods, with a $ROC\text{-}AUC$ equal to 0.94 for both EUC and DTW, outperform clearly the QDS having a $ROC\text{-}AUC$ equal to 0.86 ($TPR$\,=\,0.93 and $FPR$\,=\,0.20). The latter generates a considerable number of fake quenches, 54 in this case, as a result of some behaviors, such as digitization failures, controlled detuning and field emitters, that lead to a drop in the quality factor $Q_L$, and thus a false alarm from the QDS. For instance, the fifth station has triggered many false quenches due to a software bug, caused by a run away situation resulting from conflicting feedback algorithms.

\begin{figure}[h]
\begin{center}
\includegraphics[width=8.8cm]{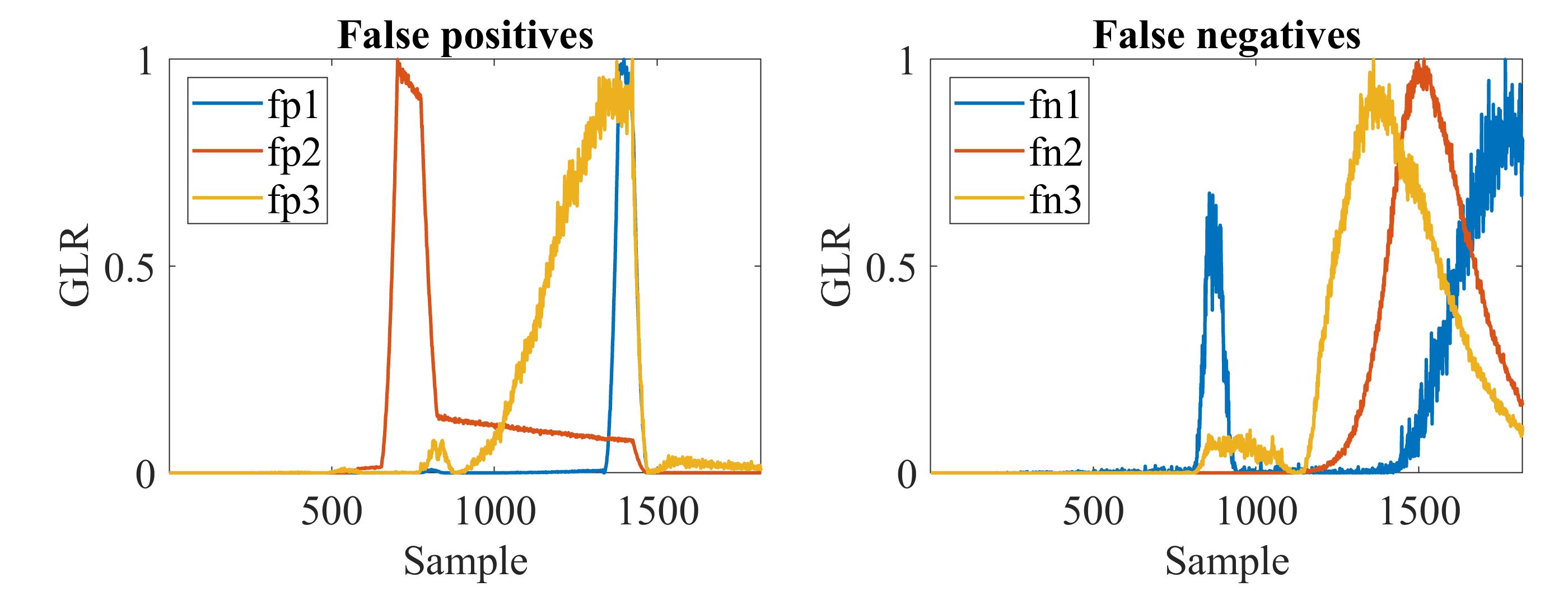}
\caption{False negative and false positive examples. fn1 is a quench trace missed by both EUC and DTW, while fn2 and fn3 were only missed by DTW. fp1 and fp2 were triggered by DTW, while fp3 was triggered by EUC.} 
\label{fn}
\end{center}
\end{figure}


Results of EUC and DTW are relatively similar, the EUC ($TPR$\,=\,0.95 and $FPR$\,=\,0.07) is however slightly better in terms of real quench detection and the DTW ($TPR$\,=\,0.93 and $FPR$\,=\,0.04) is slightly better in terms of false positives. Examples of false negatives and false positives can be seen in Fig.\ref{fn}. The false negatives are mostly caused by quenches exhibiting patterns relatively different from those used to learn the clustering models. The examples fn1 and fn2 are, for instance, caused by the beam signal, i.e., the quenches occurred simultaneously with other events, this will however be resolved upon the inclusion of the toroid. Example fn3, however, represents a quench captured at a time in between the two clusters learned with DTW. Similarly, the false positives are faults having a GLR trace with a pattern close to those of the quenches used for the model learning. The false positives fp1, fp2 and fp3 were, for instance, caused by problems with other systems than the SRFCs.

Unlike the complexity of DTW that is quadratic in the pulse length, the complexity of EUC is linear in the length of the frame of interest from the pulse. 
It's important to note that both approaches can operate on inter-pulse intervals, but using the Euclidean distance allows for intra-pulse analysis too.
Both intra-pulse and inter-pulse analyses are needed and both methods perform similarly in terms of interpretability, as both make cluster assignments mostly based on the time of the quench occurrence. The EUC therefore appears to be more advantageous for an online setup, despite the possibility of deploying both methods as part of a more collaborative strategy.

\section{Conclusion}\label{S4}
In this paper, a hybrid method to detect quenches at the EuXFEL is presented. It is a two-stage method in which the first stage generates and evaluates residuals with the nonlinear parity space method and the generalized likelihood ratio, respectively, in order to detect faults. The quenches are subsequently distinguished from the other faults in the second stage through two different clustering models, based on the EUC and DTW similarity measures. The proposed solution has been implemented and evaluated on recent data, the obtained results depict its high performance compared to the currently deployed detection system. In the near future, the beam information will be included to the implementation of the residual generation, this will help to eliminate some false alarms. The current deployment of the proposed solution is however offline, an implementation on dedicated server and on a FPGA for an online deployment has therefore been initiated. We also aim to develop and compare other detection methods and implement a human-in-the-loop procedure, in which, all methods will be deployed, and in case of disagreement, an expert intervention will be initiated. Subsequently, with the new information, there will be an automatic retraining or adjustment of the inference strategy, in order to enhance both the detection and the identification.

\bibliography{ifacconf}            
\end{document}